\documentclass[11pt,reqno,oneside]{amsart}
\usepackage{verbatim,amsmath,amssymb,cite,epsfig}
\title{Equilibrium Selection in Evolutionary Games
with Random Matching of Players} 
\author{Jacek Mi\c{e}kisz \\ Institute of Applied Mathematics \\
and Mechanics \\ Warsaw University  \\ ul. Banacha 2  \\ 02-097
Warsaw, Poland} 
\begin{document} 
\baselineskip=20pt
\maketitle 

\noindent {\bf Abstract}: 
We discuss stochastic dynamics of populations of individuals playing games. 
Our models possess two evolutionarily stable strategies: an efficient one,
where a population is in a state with the maximal payoff (fitness) 
and a risk-dominant one, where players are averse to risks. We assume 
that individuals play with randomly chosen opponents (they do not play 
against average strategies as in the standard replicator dynamics). 
We show that the long-run behavior of a population depends on its size 
and the mutation level.   
\vspace{3mm}

\vspace{3mm}

\noindent Corresponding author:  Jacek Mi\c{e}kisz, e-mail: miekisz@mimuw.edu.pl, 
phone: 48-22-5544423, fax: 48-22-5544300.
\eject

\section{Introduction}
\numberwithin{equation}{section}
\newtheorem{theo}{Theorem}
\newtheorem{defi}{Definition}
\newtheorem{hypo}{Hypothesis}

\noindent Long-run behavior of interacting individuals can be often described within 
game-theoretic models. The basic notion here is that of a Nash equilibrium. 
This is a state of population - an assignment of strategies to players - such that 
no player, for fixed strategies of his opponents, has an incentive to deviate from
his curent strategy; the change can only diminish his payoff. Nash equilibrium 
is supposed to be a result of decisions of rational players. John Maynard Smith (1974, 1982)
has refined this concept of equilibrium to include the stability of Nash equilibria against
mutants. He introduced the fundamental notion of an evolutionarily stable strategy. 
If everybody plays such a strategy, then the small number of mutants playing 
a different strategy is eliminated from the population. 
The dynamical interpretation of the evolutionarily stable strategy was later provided 
by several authors (Taylor and Jonker, 1978; Hofbauer {\em et al.}, 1979; Zeeman, 1981). 
They proposed a system of differential or difference equations, the so-called 
replicator equations, which describe the time-evolution of frequencies of strategies. 
It is known that any evolutionarily stable strategy is an asymptotically 
stable stationary point of such dynamics (Hofbauer and Sigmund, 1988; Weibull, 1995).   

Here we will discuss a stochastic adaptation dynamics of a population of players 
interacting in discrete moments of time. We will analyze two-player games 
with two strategies and two evolutionarily stable strategies. The efficient strategy
(also called payoff dominant) when played by the whole population results 
in its highest possible payoff (fitness). The risk-dominant one is played by individuals
averse to risks. The strategy is risk dominant if it has a higher expected payoff against 
a player playing both strategies with equal probabilities. We will address the problem 
of equilibrium selection - which strategy will be played in the long run with a high frequency. 

We will review two models of adaptive dynamics of a population of a fixed number of  
individuals. In both of them, the selection part of the dynamics ensures that if 
the mean payoff of a given strategy at the time $t$ is bigger than the mean payoff 
of the other one, then the number of individuals playing the given strategy should increase 
in $t+1$. In the first model, introduced by Kandori, Mailath, and Rob (1993), one assumes 
(as in the standard replicator dynamics) that individuals receive average payoffs 
with respect to all possible opponents - they play against the average strategy. 
In the second model, introduced by Robson and Vega-Redondo (1996), 
at any moment of time, individuals play only one game with randomly chosen opponents.
In both models, players may mutate with a small probability hence the population may move 
against a selection pressure. To describe the long-run behavior of such stochastic dynamics, 
Foster and Young (1990) introduced a concept of stochastic stability.
A configuration of a system is stochastically stable if it has a positive probability 
in the stationary state of the above dynamics in the limit of no mutations. 
It means that in the long run we observe it with a positive frequency. 
Kandori, Mailath, and Rob (1993) showed that in their model, 
the risk-dominant strategy is stochastically stable - 
if the mutation level is small enough we observe it in the long run 
with the frequency close to one. In the model of Robson and Vega-Redondo (1996), 
the efficient strategy is stochastically stable. 
It is one of very few models in which an efficient strategy 
is stochastically stable in the presence of a risk-dominant one. 
The population evolves in the long run to a state with the maximal fitness. 

The main goal of our paper is to investigate the effect of the number of players
on the long-run behavior of the Robson-Vega-Redondo model. We will discuss
parallel and sequential dynamics, and the one, where each individual enjoys 
each period a revision opportunity with some independent probability.   
We will show that in the last two dynamics, for any arbitrarily low 
but a fixed level of mutations, if the number of players is sufficiently big, 
a risk-dominant strategy 
is played in the long run with a frequency closed to one -  a stochastically 
stable efficient strategy is observed with a very low frequency. 
It means that when the number of players increases, the population 
undergoes a transition between an efficient payoff-dominant equilibrium 
and a risk-dominant one. We will also show that for some range of payoff parameters,
stochastic stability itself depends on the number of players. If the number of players 
is below certain value (which may be arbitrarily large), then a risk-dominant strategy 
is stochastically stable. Only if $n$ is large enough, an efficient strategy 
becomes stochastically stable as proved by Robson and Vega-Redondo (1996).

In Section 2, we introduce Kandori-Mailath-Rob and Robson-Vega-Redondo models
and review their main properties. In Section 3, we analyze the Robson-Vega-Redondo model
in the limit of the infinite number of players and show our main results. 
Discussion follows in Section 4. 

\section{Models of Adaptive Dynamics with Mutations}

\noindent We will consider a finite population of $n$ individuals
who have at their disposal one of two strategies:
$A$ and $B$. At every discrete moment of time, $t=1,2,...,$
they are randomly paired (we assume that $n$ is even)
to play a two-player symmetric game with payoffs given by the following matrix:
\vspace{5mm}

\hspace{23mm} A \hspace{2mm} B   

\hspace{15mm} A \hspace{3mm} a \hspace{3mm} b 

U = \hspace{6mm} 

\hspace{15mm} B \hspace{3mm} c \hspace{3mm} d

where the $ij$ entry, $i,j = A, B$, is the payoff of the first (row) player when
he plays the strategy $i$ and the second (column) player plays the strategy $j$. 
We assume that both players are the same and hence payoffs of the column player are given 
by the matrix transposed to $U$; such games are called symmetric. 

We assume that $a>c$ and $d>b$, therefore both $A$ and $B$ 
are evolutionarily stable strategies,
and $a+b<c+d$, so the strategy $B$ has a higher expected payoff against a player playing 
both strategies with the probability $1/2$. We say that $B$ risk dominates the strategy $A$ 
(the notion of the risk-dominance was introduced and thoroughly studied 
by Hars\'{a}nyi and Selten (1988)). We also assume that $a>d$ hence 
we have a selection problem of choosing between the risk-dominant $B$ 
and the so-called payoff-dominant or efficient strategy $A$.

At every discrete moment of time $t$, the state of our population
is described by the number of individuals, $z_{t}$, playing $A$. 
Formally, by the state space we mean the set
$$\Omega=\{z, 0 \leq z\leq n\}.$$
Now we will describe the dynamics of our system.
It consists of two components: selection and mutation.
The selection mechanism ensures that if the mean payoff 
of a given strategy, $\pi_{i}(z_{t}), i=A,B$, 
at the time $t$ is bigger than the mean payoff 
of the other one, then the number of individuals
playing the given strategy should increase in $t+1$. 
In their paper, Kandori, Mailath, and Rob (1993) write
\begin{equation}
\pi_{A}(z_{t})=\frac{a(z_{t}-1)+b(n-z_{t})}{n-1},
\end{equation} 
$$\pi_{B}(z_{t})=\frac{cz_{t}+d(n-z_{t}-1)}{n-1},$$
provided $0<z_{t}<n$.

It means that in every time step, players are paired infnitely many times
to play the game or equivalently, each player plays with every other player 
and his payoff is the sum of corresponding payoffs. This model may be therefore
considered as an analog of replicator dynamics for populations with fixed numbers 
of players.

The selection dynamics is formalized in the following way:

\begin{equation}
z_{t+1} > z_{t} \hspace{2mm} if \hspace{2mm} \pi_{A}(z_{t}) > \pi_{B}(z_{t}),
\end{equation}
$$z_{t+1} < z_{t} \hspace{2mm} if \hspace{2mm} \pi_{A}(z_{t}) < \pi_{B}(z_{t}),$$
$$z_{t+1}= z_{t} \hspace{2mm} if \hspace{2mm} \pi_{A}(z_{t}) = \pi_{B}(z_{t}),$$
$$z_{t+1}= z_{t} \hspace{2mm} if \hspace{2mm} z_{t}=0  \hspace{2mm} or \hspace{2mm} z_{t}=n.$$

Now mutations are added. Players may switch to new strategies with the probability $\epsilon$. 
It is easy to see that for any two states of the population there is a positive probability
of the transition between them in some finite number of time steps. 
We have therefore obtained an irreducible Markov chain with $n+1$ states. 
It has a unique stationary probability distribution 
(a stationary state) which we denote by $\mu^{\epsilon}_{n}.$
It was shown (Kandori {\em et al.}, 1993) 
that $\lim_{\epsilon \rightarrow 0}\mu^{\epsilon}_{n}(0)=1$
which means that in the long run, in the limit of no mutations, all players play
the risk-dominant strategy $B$. We say that the risk-dominant strategy 
is {\em stochastically stable}. 

The general set up in the Robson-Vega-Redondo model (1996) is the same.
However, individuals are paired only once at every time step and play only one game
before selection process takes place. Let $p_{t}$ denote the random variable
which describes the number of cross-pairings, i.e. the number of pairs 
of matched individuals playing different strategies at the time $t$.
Let us notice that $p_{t}$ depends on $z_{t}$.
For a given realization of  $p_{t}$ and $z_{t}$,
mean payoffs obtained by each strategy are as follows:
\begin{equation}
\tilde{\pi}_{A}(z_{t},p_{t})=\frac{a(z_{t}-p_{t})+bp_{t}}{z_{t}},
\end{equation} 
$$\tilde{\pi}_{B}(z_{t},p_{t})=\frac{cp_{t}+d(n-z_{t}-p_{t})}{n-z_{t}},$$
provided $0<z_{t}<n$. Then the authors show that the payoff-dominant strategy 
is stochastically stable. We will outline their proof.

First of all, one can show that there exists $k$ such that 
if $n$ is large enough and $z_{t} \geq k$, 
then there is a positive probability (a certain realization of $p_{t}$)
that after a finite number of steps of the mutation-free selection dynamics,
all players will play $A$. Likewise, if $z_{t} <k$ (for any $k \geq 1$), 
then if the number of players 
is large enough, then after a finite number of steps of the mutation-free selection dynamics 
all players will play $B$. In other words, $z=0$ and $z=n$ 
are the only absorbing states of the mutation-free dynamics.
Moreover, if $n$ is large enough, then if $z_{t} \geq n-k$, then the mean payoff obtained
by $A$ is always (for any realization of $p_{t}$) bigger than the mean payoff obtained by $B$
(in the worst case all $B$-players play with $A$-players). Therefore the size of the basin 
of attraction of the state $z=0$ is at most $n-k-1$ and that of $z=n$ is at least $n-k$. 
Observe that mutation-free dynamics is not deterministic ($p_{t}$ describes the random matching)
and therefore basins of attraction may overlap. It follows that the system needs at least $k+1$ 
mutations to evolve from $z=n$ to $z=0$ and at most $k$ mutations to evolve from $z=0$ to $z=n$. 
Now using a tree representation of stationary states of irreducible Markov chains 
(Freidlin and Wentzell, 1970 and 1984; see also Appendix B) Robson and Vega-Redondo finish the proof 
and show that the efficient strategy is stochastically stable. 

However, as outlined above, their proof requires the number of players to be sufficiently large. 
We will now show that a risk-dominant strategy is stochastically stable 
if the number of players is below certain value which can be arbitrarily big.

If the population consists of only one $B$-player and $n-1$ $A$-players 
and if $c>[a(n-2)+b]/(n-1)$, that is $n< (2a-c-b)/(a-c)$, 
then  $\tilde{\pi}_{B}> \tilde{\pi}_{A}.$ It means that one needs only one mutation 
to evolve from $z=n$ to $z=0.$ It is easy to see that two mutations are necessary 
to evolve from $z=0$ to $z=n.$ Using again the tree representation of stationary states 
one can prove the following theorem.

\begin{theo}
If $n<\frac{2a-c-b}{a-c}$, then the risk-dominant strategy $B$ is stochastically stable
in the case of random matching of players.
\end{theo}      

To see stochastically stable states, we need to take the limit of no mutations. 
We will now examine the long-run behavior of the Robson-Vega-Redondo model 
for a fixed level of mutations in the limit of the infinite number of players.

\section{Long-Run Behavior in the Limit of Infinitely Many Players}

\noindent We will consider three specific cases of the selection rule (2.2).

In the parallel dynamics, everyone in the selection process chooses 
at the same time (all players are synchronized) a strategy with 
the bigger average payoff. It means that after mutations have taken place, 
the selection moves the population to one of the two extreme states, 
$z=0$ or $z=n$. Our system becomes then a two-state Markov chain
with a unique stationary state $\mu_{n}^{\epsilon}$ 
(a similar model was discussed in (Vega-Redondo, 1997)). 
We will show that for any number of players, if the mutation level 
is sufficiently small, then in the long run almost all individuals play 
the payoff-dominant strategy. The same result holds for any small mutation level 
if the number of players is large enough. 
\vspace{2mm}

\noindent {\bf Theorem 2}
\vspace{2mm}
 
\noindent In the parallel dynamics, 
\vspace{2mm}

$$\lim_{\epsilon \rightarrow 0}\mu_{n}^{\epsilon}(n)=1 \; \; for \; \; every \; \; n,$$
$$\lim_{n \rightarrow \infty}\mu_{n}^{\epsilon}(n)=1 \; \; for \; \; every \; \;  \epsilon <1 .$$
\vspace{3mm}

{\bf Proof:} We are looking for a unique stationary state $\mu_{n}^{\epsilon}$ 
of a two-state Markov chain. Let us denote by $p_{0n}$ a transition probability
from the state $z=0$ to $z=n$ and by $p_{n0}$ from $z=n$ to $z=0$.
We have
\begin{equation}
\mu_{n}^{\epsilon}(n)= \frac{p_{0n}}{p_{0n}+p_{n0}}.
\end{equation}
For the transition from $z=0$ to $z=n$ it is enough that two players mutate 
from $B$ to $A$ and then they are paired to play a game. It follows that
\begin{equation}
p_{0n} > \epsilon^{2}\frac{1}{n-1}.
\end{equation}
Transition from $z=n$ to $z=0$ requires at least $\gamma n$ mutations 
(for some $\gamma$) which means that
\begin{equation}
p_{n0} < \epsilon^{\gamma n}
\end{equation} 
It follows from (3.1-3.3) that 
\begin{equation}
\mu_{n}^{\epsilon}(n) >\frac{1}{1+(n-1)\epsilon^{\gamma n-2}}
\end{equation}
Hence $\mu_{n}^{\epsilon}(n)$ is arbitrarily close to one
if $\epsilon$ is sufficiently small or $n$ is sufficiently big.  
\vspace{3mm}

Now we will analyze the other extreme case of a selection rule (2.2) - 
a sequential dynamics where in one time unit only one player can change his strategy. 
Although our dynamics is discrete in time, it captures the essential features
of continuous-time models, where every player has an exponentially
distributed waiting time to a moment of a revision opportunity. 
Probability that two or more players revise their strategies 
at the same time is therefore equal to zero - this is an example 
of a birth and death process. 

The number of $A$-players in the population may increase by one
in $t+1$, if a $B$-player is chosen in $t$ which happens with 
the probability $(n-z_{t})/n$. Analogously, the number of $B$-players 
in the population may increase by one in $t+1$, if an $A$-player 
is chosen in $t$ which happens with the probability $(z_{t})/n$. 

The player who has a revision opportunity chooses in $t+1$ 
with the probability $1-\epsilon$ the strategy with a higher average payoff in $t$ 
and the other one with the probability $\epsilon$.  

Let $r(k)=P(\tilde{\pi}_{A}(z_{t},p_{t}) > \tilde{\pi}_{B}(z_{t},p_{t}))$
and $l(k)=P(\tilde{\pi}_{A}(z_{t},p_{t}) < \tilde{\pi}_{B}(z_{t},p_{t}))$.
The sequential dynamics is described by the following transition probabilities:

if $z_{t}=0 $, then $z_{t+1}=1 $ with the probability $\epsilon$
and $z_{t+1}=0 $ with the probability $1-\epsilon$,

if $z_{t}=n $, then $z_{t+1}=n-1 $ with the probability $\epsilon$
and $z_{t+1}=n $ with the probability $1-\epsilon$,

if $z_{t} \neq 0,n$, then $z_{t+1}= z_{t} +1$ with the probability

$$r(k)\frac{n-z_{t}}{n}(1-\epsilon)+(1-r(k))\frac{n-z_{t}}{n}\epsilon$$

and $z_{t+1}= z_{t} -1$ with the probability

$$l(k)\frac{z_{t}}{n}(1-\epsilon)+(1-l(k))\frac{z_{t}}{n}\epsilon.$$

In the dynamics intermediate between the parallel and sequential one, at time period, 
each individual has a revision opportunity with some probability $\tau <1$. 
Each chosen player follows independently the same rule as in the sequential dynamics.
The probability that in one period, a given player will have a revision opportunity
should be proportional to the length of the period (which we normalized to $1$ in our models).
For a fixed $\epsilon$ and an arbitrarily large but fixed $n$, we consider the limit of continuous time, 
$\tau \rightarrow 0$, and show that the limiting behavior 
is already obtained for a sufficiently small $\tau$, namely  $\tau  < \epsilon /n^{3}$. 
 
For an interesting discussion on the importance of the order of taking different limits 
$(\tau  \rightarrow 0, n \rightarrow \infty,$ and $\epsilon \rightarrow 0)$
in evolutionary models (especially in the Aspiration and Imitation model) 
see Samuelson (1997).

In the intermediate dynamics, instead of $(n-z_{t})/n$ and 
$z_{t}/n$ probabilities we have more involved combinatorial 
factors. In order to get rid of these inconvenient factors, we will enlarge the state space
of the population. The state space $\Omega^{'}$ is the set of all configurations 
of players, that is all possible assignments of strategies to individual players.
Therefore, a state $z_{t}=k$ in $\Omega$ consists of 
$\left( \begin{array}{c} n \\ k \end{array} \right)$ 
states in $\Omega^{'}$. The sequential dynamics is not anymore 
a birth and death process on $\Omega^{'}$. 
However, we will be able to treat both dynamics in the same framework.

We will show that for any arbitrarily low but fixed level of mutation, 
if the number of players is large enough, then in the long run only 
a small fraction of the population play the payoff-dominant strategy. 
Smaller the mutation level is, fewer players use the payoff-dominant strategy.  

The following two theorems are proved in Appendix C. 

\noindent {\bf Theorem 3}
\vspace{2mm}

\noindent In the sequential dynamics, for any $\delta >0$ 
and $\beta >0$ there exist $\epsilon(\delta, \beta)$
and $n(\epsilon)$ such that for any $n > n(\epsilon)$
$$\mu_{n}^{\epsilon}(z \leq \beta n) > 1- \delta.$$
\vspace{2mm}

\noindent {\bf Theorem 4}
\vspace{2mm}

\noindent In the intermediate dynamics dynamics, for any $\delta >0$ 
and $\beta >0$ there exist $\epsilon(\delta, \beta)$
and $n(\epsilon)$ such that for any $n > n(\epsilon)$ and $\tau< \frac{\epsilon}{n^{3}}$
$$\mu_{n}^{\epsilon}(z \leq \beta n) > 1- \delta.$$
\vspace{2mm}

Let us note that the above theorems concern an ensemble of configurations,
not an individual one. In the limit of the infinite number of players, that is 
the infinite number of configurations, every single configuration has zero probability
in the stationary state. It is an ensemble of configurations that might be stable
(Mi\c{e}kisz, 2004a and 2004b).  

Let us now assume that at every time period, players are matched many times. 
It follows from the results in (Kandori {\em et al.}, 1993; Robson and Vega-Redondo, 1996) 
analysed in (Vega-Redondo, 1996) that the limits of zero mutations 
and the infinite number of matching rounds per period do not commute. 
In the limit of the infinite number of matching rounds per period, 
individuals play against the average strategy and we obtain 
the Kandori-Mailath-Rob model and their conclusion follows. 
On the other hand, for any fixed number of matching rounds (the Robson-Vega-Redondo model), 
the limit of zero mutations gives us the stochastic stability 
of an efficient strategy. Here we investigated the effect of the number 
of players on the long-run behavior in the random matching model. 
We showed that the limit of the infinite number of players has the same effect as the limit
of the infinite number of matching rounds. In fact, the probability that the average payoff 
of strategy $A$ is bigger than the average payoff of strategy $B$ converges
in both limits to $1$ or $0$, if the fraction of the population playing $A$ 
is respectively right to or left to the unique mixed Nash equilibrium. 
Both limits are therefore alternative ways of representing the idea 
of a low matching-induced noise. 

\section{Conclusion}

\noindent We studied the effect of the number of players on the long-run behavior 
in the adaptive dynamics with mutations and random matching of players. 
We showed that in the sequential dynamics for any arbitrarily low 
but fixed level of mutation, if the number of players is large enough, 
then in the long run almost all of them play a risk-dominant strategy. 
The same result holds if at any period, each individual has a revision opportunity 
with some small probability. This is in contrast with the result of Robson 
and Vega-Redondo (1996) who for a fixed number of players take the limit of zero mutations 
and obtain stochastic stability of a payoff-dominant strategy. It means that when 
the number of players increases, the population undergoes a transition between 
an efficient payoff-dominant equilibrium and a risk-dominant one. Therefore, 
in any specific model, to describe its long-run behavior, one has to evaluate 
the number of players and the mutation level. 
\vspace{3mm}

\noindent {\bf Acknowledgments}: I would like to thank the Polish Committee 
for Scientific Research for financial support under the grant KBN 5 P03A 025 20.

\appendix

\section{}

\noindent {\bf Random variable of cross-pairings}
\vspace{3mm}

\noindent We will first investigate the random variable $p_{t}$ which describes
the number of cross-pairings in a state $z_{t}$. Let $z_{t} = \alpha n$.
Let $P$ be the probability mass function of the random variable 
of cross-pairings $p$; we skip the subscript $t$.

\noindent {\bf Proposition 1}
\begin{equation} 
P(|p-n\alpha(1-\alpha)|>\beta n) \rightarrow_{n \rightarrow \infty} 0
\end{equation}
for any $\beta>0$.
\vspace{2mm}

{\bf Proof:} Let the number of A-players be equal to $k=\alpha n$.
We begin by dividing all players into two groups. We arrange them randomly in a row
and pick the first $n/2$ ones to be members of the first group. 
This will be players who will choose randomly their opponents. 
Let $X$ denote the random variable counting the number of A-players in this group,
$X= X_{1}+...+X_{n/2}$, where $X_{i}=1$ if the i-th player plays $A$; otherwise $X_{i}=0$.
The expected value and the variance of $X_{i}$ are equal to 
$E(X_{i})=\alpha$ and $Var(X_{i})=\alpha(1-\alpha)$ respectively.
One then have that 
\begin{equation}
E(X)=E(X_{i})+...+E(X_{n/2})=\alpha n/2.
\end{equation}
\begin{equation}
Var(X)=Var(X_{i})+...+Var(X_{n/2})+2\sum_{j<k}(E(X_{j}X_{k})-E(X_{j})E(X_{k}))=
\frac{n}{2}\alpha(1-\alpha)(1-\frac{n-2}{2(n-1)}).
\end{equation}

From the Czebyshev inequality we get that 
\begin{equation}
P(|X-E(X)|>\beta_{1} n) \leq \frac{Var(X)}{(\beta_{1} n)^{2}}\rightarrow_{n \rightarrow \infty} 0.
\end{equation}
for every $\beta_{1} >0$.

Now every player from the first group is randomly paired with a player
from the second group. Let us first assume (for pedagogical reasons)
that the number of A-players in the first group (and therefore in the second group) 
is exactly equal to $\alpha n/2$.

Let $Y$ be the random variable describing 
the number of cross-pairings for a given realization of $X$. 
$Y=Y_{i}+...+Y_{n/2}$, where $Y_{i}=1$
if the i-th player has chosen the opponent with a different strategy; 
$Y_{i}=0$ otherwise.
The expected value of $Y_{i}$ is equal to 
$E(Y_{i})=1-\alpha$ if $X_{i}=1$ and $E(Y_{i})=\alpha$ if $X_{i}=0$, 
and $Var(Y_{i})=\alpha(1-\alpha)$.

We get 
\begin{equation}
E(Y)=E(Y_{i})+...+E(Y_{n/2})=\alpha(1-\alpha)n
\end{equation}
and using the formula for the variance of the sum of the random variables in (A.3) we obtain
\begin{equation}
Var(Y)=\frac{n}{2}\alpha(1-\alpha)+\frac{n\alpha(1-\alpha)}{n-2}+
\frac{n^{2}\alpha(1-\alpha)-(3\alpha^{2}-3\alpha+1)}{2n-4}.
\end{equation}

Now let the number of A-players in the first group be equal to $(\alpha+\alpha_{1})n/2$
for some $\alpha_{1}$. We get that
\begin{equation}
E(Y)=n\alpha(1-\alpha)+n\alpha_{1}^{2}.
\end{equation}
We again use the formula in (A.3) and get that
\begin{equation}
Var(Y)=C(\alpha,\alpha_{1})O(n),
\end{equation}
where $C(\alpha,\alpha_{1})$ is some constant depending on $\alpha$ and $\alpha_{1}$
and $lim_{n \rightarrow \infty}O(n)/n =1$.  
For any fixed number of A-players in the first group we use 
the Czebyshev inequality to get 
\begin{equation}
P(|Y-E(Y)|> \beta_{2} n) \rightarrow_{n \rightarrow \infty} 0.
\end{equation}
for every $\beta_{2} >0$.

Now we set $\beta_{1} = \alpha_{1}^{2}$ in (A.4).
Then Proposition 1 follows with $\beta = \beta_{1}+\beta_{2}$.
\vspace{3mm}

Now for any state of the system, $z=k,k \neq 0, n$, 
we will calculate, in the limit of the infinite number
of players, the probability, $r(k)$, that the average payoff of $A$ 
is bigger than that of $B$. We have
\begin{equation}
r(k)=P(\frac{a(k-p)+bp}{k} > \frac{cp+d(n-k-p)}{n-k}).
\end{equation}

Let $k= \alpha n$. It follows from (A.10) that
\begin{equation}
r(\alpha n)=P(p(\frac{d-c}{n(1-\alpha)}+\frac{b-a}{\alpha n}) >d-a).
\end{equation}
If $(d-c)/(1-\alpha)+(b-a)/\alpha \geq 0$, then $r(\alpha n)=1$ because $d<a$.
This happens for 
$\alpha \geq (a-b)/(a-c+d-b) \equiv \gamma_{1}>1/2$. Let us notice that 
if $c \leq d$, then $\gamma_{1} \leq 1$, if $c>d$, then $\gamma_{1} >1$.
For $\alpha <\gamma_{1}$, from (A.11) we get 
\begin{equation}
r(\alpha n)=P(p < \frac{n(d-a)(1-\alpha)\alpha}{(d-c)\alpha+(b-a)(1-\alpha)}).              
\end{equation}

Now it follows from Proposition 1 that if 
\begin{equation}
\frac{(d-a)(1-\alpha)\alpha}{(d-c)\alpha+(b-a)(1-\alpha)} <\alpha(1-\alpha),
\end{equation}
which holds for  $\alpha < \gamma_{2} \equiv \frac{d-b}{d-b+a-c}$, then
\begin{equation}
lim_{n \rightarrow \infty}r(\alpha n) =0 .
\end{equation}
Note that $\gamma_{2}$, ($1/2 < \gamma_{2} < 1$), is the unique mixed Nash equilibrium 
of the game. We have proved the following proposition.
\vspace{2mm}

\noindent {\bf Proposition 2}

If $\alpha \geq \gamma_{1}$, then $r(\alpha n) = 1$,
 
if $\gamma_{2} <\alpha < \gamma_{1}$, then $lim_{n \rightarrow \infty}r(\alpha n) = 1$,
 
if $\alpha < \gamma_{2}$, then $lim_{n \rightarrow \infty}r(\alpha n) = 0.$ 

\section{}

\noindent {\bf Stationary states of irreducible Markov chains}
\vspace{3mm}

\noindent The following tree representation of stationary states 
of Markov chains was proposed by Freidlin and Wentzell (1970 and 1984). 
Let $(\Omega,P)$ be an irreducible Markov chain with a state space 
$\Omega$ and transition probabilities given by $P: \Omega \times \Omega \rightarrow [0,1]$. 
It has a unique stationary probability distribution $\mu$ (called also a stationary state). 
For $x \in \Omega$, an $x$-tree is a directed graph 
on $\Omega$ such that from every $y \neq x$ there is a unique path to $x$
and there are no outcoming edges out of $x$. Denote by $T(x)$ the set of all $x$-trees
and let 
\begin{equation}
q(x)=\sum_{d \in T(x)} \prod_{(y,y') \in d}P(y,y'),
\end{equation}
where the product is with respect to all edges of $d$. 
Now one can show that
\begin{equation}
\mu(x)=\frac{q(x)}{\sum_{y \in \Omega}q(y)}
\end{equation}
for all $x \in \Omega.$

A state is an absorbing one if it attracts nearby states in the mutation-free dynamics. 
We assume that after a finite number of steps of the mutation-free dynamics 
we arrive at one of the absorbing states (there are no other recurrence classes)
and stay there forever. Then it follows from the above tree representation 
that any state different from absorbing states 
has zero probability in the stationary distribution in the zero-mutation limit. 
Moreover, in order to study the zero-mutation limit of the stationary state, 
it is enough to consider paths between absorbing states. More precisely, 
we construct $x$-trees with absorbing states as vertices; the family of such 
$x$-trees is denoted by $\tilde{T}(x)$. Let 
\begin{equation}
q_{m}(x)=max_{d \in \tilde{T}(x)} \prod_{(y,y') \in d}\tilde{P}(y,y'),
\end{equation}
where $\tilde{P}(y,y')= max \prod_{(w,w')}P(w,w')$,
where the product is taken along any path joining $y$ with $y'$ and the maximum 
is taken with respect to all such paths. 
Now we may observe that if $lim_{\epsilon \rightarrow 0} q_{m}(y)/q_{m}(x)=0,$
for any $y\neq x$, then $x$ is stochastically stable. Therefore we have to compare  
trees with the biggest products in (B.3); such trees we call maximal.
\newpage

\section{}

\noindent {\bf Proof of Theorem 3}
\vspace{2mm}

\noindent Pick $\delta, \beta$, and $\epsilon$.
It follows from the limiting properties of $r(k)$ that there is $n(\epsilon, \delta)$
and $1/2< \gamma_{3} < \gamma_{2}$ such that for all $n>n(\epsilon, \delta)$ 
we have that $r(\alpha n)< \epsilon$ if $\alpha \leq \gamma_{3}$.

For any state in $\Omega^{'}$ with $z=k$, we will prove that 
\begin{equation}
q(k) < 3\epsilon q(k-1), \hspace{3mm} 1 \leq k  \leq  \gamma_{3}n,
\end{equation}
\begin{equation}
q(k) < \frac{2q(k-1)}{\epsilon}, \hspace{3mm} \gamma_{3}n < k \leq n.
\end{equation}

It follows from (C.1-C.2) that

\begin{equation}
\mu^{\epsilon}_{n}(z \leq \beta n)=\frac{\sum_{0 \leq k \leq \beta n} 
\left( \begin{array}{c} n \\ k \end{array} \right) q(k)}
{\sum_{0 \leq k \leq \beta n}
\left( \begin{array}{c} n \\ k \end{array} \right)q(k)+
\sum_{k > \beta n}^{k \leq \gamma_{3}n}
\left( \begin{array}{c} n \\ k \end{array} \right) q(k)+
\sum_{k > \gamma_{3}n}^{n}
\left( \begin{array}{c} n \\ k \end{array} \right)q(k)}
\end{equation} 

$$> \frac{1}{1+\sum_{k > \beta n}^{k \leq \gamma_{3}n}
\left( \begin{array}{c} n \\ k \end{array} \right)(3\epsilon)^{k}+
(3\epsilon)^{\gamma_{3}n-1} \sum_{k > \gamma_{3}n}^{n}
\left( \begin{array}{c} n \\ k \end{array} \right)
(\frac{2}{\epsilon})^{k-\gamma_{3}n}}$$

$$> \frac{1}{1+\sum_{k > \beta n}^{k \leq \gamma_{3}n}(3\epsilon)^{k}
e^{k}(\frac{n}{k})^{k}+ 
\epsilon^{(2\gamma_{3}-1)n-1}3^{n-1}(\frac{\epsilon}{3}+1)^{n}} >1-\delta$$

if $\epsilon$ is small enough. Smaller $\beta$ is and $\gamma_{3}$ closer
to $1/2$, smaller $\epsilon$ should be.

To prove (C.1-C.2), with every $k$-tree ($1 \leq k \leq \gamma_{3}n$) 
we will associate a ($k-1$)-tree. Let $\omega$ be a $k$-tree. 
We reverse arrows on all edges on the unique path between $k-1$ and $k$ 
(all other edges we leave unchanged). (C.1) follows from the bound
$$\frac{r(k)(1-\epsilon)+(1-r(k))\epsilon}{(1-r(k-1))(1-\epsilon)+r(k-1)\epsilon}<3\epsilon$$
and (C.2) from the bound
$$\frac{r(k)(1-\epsilon)+(1-r(k))\epsilon}{(1-r(k-1))(1-\epsilon)+r(k-1)\epsilon}
>\frac{\epsilon}{2}.$$
\vspace{2mm}

\noindent {\bf Proof of Theorem 4}
\vspace{2mm}

In the intermediate dynamics, the probability of moving $m$ units
to the right if $r(k)<\epsilon$ or to the left if $1-r(k) < \epsilon$ 
is not proportional to $\epsilon^{m}$ as in the sequential dynamics.
Therefore to prove (C.1-C.2) we cannot simply 
reverse arrows on edges in constructing corresponding trees.

To prove (C.1), with every $k$-tree ($1 \leq k \leq \gamma_{3}n$) 
we will again associate a ($k-1$)-tree. Let $\omega$ be a $k$-tree. 
If on the unique path between $k-1$ and $k$ there are only transitions 
which involve only one individual at any time period, then we reverse 
arrows on all edges on this path as in the proof of Theorem 3. 
Otherwise, let an edge $j \rightarrow l$ be the first edge which involves at least two players.  
If $j > k-1$, then we reverse all arrows between $k-1$ and $j$, 
cut the edge $j \rightarrow l$ and connect $k$ to $k-1$.
Because an edge was deleted, a correspondence between $k$ 
and ($k-1$)-trees is not one-to-one anymore. If the edge $j \rightarrow l$
involves $m$ players, then there are at most 
$\left( \begin{array}{c} n \\ m \end{array} \right) $
$k$-trees with the same corresponding ($k-1$)-tree.
By cutting the edge we decreased a probability at least $\tau^{m}$ times.
If $\tau n^{2} < 1/2$, then  the series 
$\sum_{m \geq 2}\tau^{m}\left( \begin{array}{c} n \\ m \end{array} \right)$
is bounded by $\tau$. C.1 follows.

If $j \leq k-1$, then we cut the edges $k-1 \rightarrow $
and $j \rightarrow l$, connect $j$ to a state with $z=j-1$
(only one player changes his strategy) and $k$ to $k-1$.
By the above procedure we decreased a probability by 
$\tau$. There are at most $n^{3}$ $k$-trees 
with the same corresponding ($k-1$)-tree. 
If $ \tau n^{3} < \epsilon$, then (C.1) follows.
 
(C.2) can be proved in an analogous way. Now Theorem 4 follows 
in the same way as Theorem 3.

\noindent {\bf References}
\vspace{3mm}

\noindent Foster, D. and Young, P. H., 1990.
Stochastic evolutionary game dynamics.
{\em Theoretical Population Biology} 38: 219-232.
\vspace{2mm}

\noindent Freidlin, M. and Wentzell, A., 1970.
On small random perturbations of dynamical systems. {\em Russian Math. Surveys} 25: 1-55.
\vspace{2mm}

\noindent Freidlin, M. and Wentzell, A., 1984. {\em Random Perturbations
of Dynamical Systems.} Springer Verlag, New York.
\vspace{2mm}

\noindent Hars\'{a}nyi, J. and Selten, R., 1988.
{\em A General Theory of Equilibrium Selection in Games.} 
MIT Press, Cambridge MA. 
\vspace{2mm}

\noindent Hofbauer, J., Schuster, P., and Sigmund, K., 1979.
A note on evolutionarily stable strategies and game dynamics.
{\em J. Theor. Biol.} 81: 609-612.
\vspace{2mm}

\noindent Hofbauer, J. and Sigmund, K., 1988. {\em The Theory
of Evolution and Dynamical Systems.} Cambridge University Press, Cambridge.
\vspace{2mm}

\noindent Kandori, M., Mailath, G. J., and Rob, R., 1993.
Learning, mutation, and long-run equilibria in games.
{\em Econometrica} 61: 29-56.
\vspace{2mm}

\noindent Maynard Smith, J., 1974. The theory of games and the evolution
of animal conflicts. {\em J. Theor. Biol.} 47: 209-221.
\vspace{2mm}

\noindent Maynard Smith, J., 1982. {\em Evolution and the Theory of Games}.
Cambridge University Press, Cambridge. 
\vspace{2mm}

\noindent Mi\c{e}kisz, J., 2004a. Stochastic stability of spatial three-player games.
Warsaw University preprint, www.mimuw.edu.pl/$\sim$miekisz/physica.ps, 
to appear in {\em Physica A}.
\vspace{2mm}

\noindent Mi\c{e}kisz, J., 2004b. Statistical mechanics of spatial evolutionary games.
Warsaw University preprint, www.mimuw.edu.pl/$\sim$miekisz/statmech.ps, 
to appear in {\em J. Phys. A}.
\vspace{2mm}

\noindent Robson, A. and Vega-Redondo, F., 1996. Efficient equilibrium selection
in evolutionary games with random matching. {\em J. Econ. Theory} 70: 65-92.
\vspace{2mm}

\noindent Samuelson, L., 1997. {\em Evolutionary Games and Equilibrium Selection}.
MIT Press, Cambridge MA.
\vspace{2mm}

\noindent Taylor, P. D. and Jonker, L. B., 1978. Evolutionarily stable
strategy and game dynamics. {\em Math. Biosci.} 40: 145-156.
\vspace{2mm}

\noindent Vega-Redondo, F., 1996. {\em Evolution, Games, and Economic Behaviour.}
Oxford University Press, Oxford.
\vspace{2mm}

\noindent Weibull, J., 1995. {\em Evolutionary Game Theory.} 
MIT Press, Cambridge MA.
\vspace{2mm}

\noindent Zeeman, E., 1981. Dynamics of the evolution of animal conflicts.
{\em J. Theor. Biol.} 89: 249-270.

\end{document}